\documentclass[a4paper,11pt]{article}
\pdfoutput=1 

\usepackage{jcappub} 

\usepackage[T1]{fontenc} 

\usepackage{mathptmx}
\usepackage{verbatim}
\usepackage{multirow}
\usepackage{cuted}
\usepackage{graphicx}

\RequirePackage{graphicx}
\RequirePackage{amsmath}
\RequirePackage[cal=cm,scr=boondoxo,scrscaled=1.05]{mathalfa}  

\usepackage{placeins}

\newcommand{\ee}{\mathrm{e}}  
\newcommand{\rs}{r_\mathrm{s}}
\newcommand{\dm}{\mathrm{\Delta}M}
\newcommand{\drs}{\mathrm{\Delta}\rs}
\newcommand{\cl}{\mathcal{L}}
\newcommand{\ce}{\mathcal{E}}

\newcommand{\diff}{\mathrm{d}}

\title{
Supermassive black holes surrounded by dark matter modeled as anisotropic fluid: epicyclic oscillations and their fitting to observed QPOs}

\author[a]{Z. Stuchl\'{\i}k,}
\author[a,1]{and J. Vrba \note{Corresponding author.}}

\affiliation[a]{Research Centre for Theoretical Physics and Astrophysics, Institute of Physics, Silesian University in Opava, Bezru\v{c}ovo n\'am.~13, 746\,01, Opava, CZ}

\emailAdd{zdenek.stuchlik@physics.slu.cz}
\emailAdd{jaroslav.vrba@physics.slu.cz}

\abstract{Recently introduced exact solution of the Einstein gravity coupled minimally to an anisotropic fluid representing dark matter can well represent supermassive black holes in galactic nuclei with realistic distribution of dark matter around the black hole, given by the Hernquist-like density distribution. For these fluid-hairy black hole spacetimes, properties of the gravitational radiation, quasinormal ringing, and optical phenomena were studied, giving interesting results. Here, using the range of physical parameters of these spacetimes allowing for their relevance in astrophysics, we study the epicyclic oscillatory motion of test particles in these spacetimes. The frequencies of the orbital and epicyclic motion are applied in the epicyclic resonance variant of the geodesic model of quasiperiodic oscillations (QPOs) observed in active galactic nuclei to demonstrate the possibility to solve the cases where the standard vacuum black hole spacetimes are not allowing for explanation of the observed data. We demonstrate that the geodesic model can explain the QPOs observed in most of the active galactic nuclei for the fluid-hairy black holes with reasonable halo parameters.}

\begin{document}
\maketitle
\flushbottom

\section{Introduction}
The dark energy and dark matter naturally play crucial role in cosmology \cite{Linde:1990:PPIC:,Kra-Tur:1995:GRG:,Bah-Ost-Per:1999:Sci:,Rie-etal:2004:APJ:,Cal-Kam:2009:Nat:,Ada-Dur-Gue:2013:ASTRA:,Ade-Pla-Col:2014:ASTRA:}, but they are very important also in astrophysics, namely for structure of galaxies and their clusters \cite{Stu:2005:MPLA:,Stu-Hle-Nov:2016:PRD:,Stu-etal:2020:Uni:}, or the motion around large galaxies \cite{Stu-Sche:2011:JCAP:,Stu-Char-Sche:2018:EPJC:}. 

The observational data indicate that the dark energy is well represented by the relict cosmological constant, having strongest influence in the edge of large galaxies that nearly coincides with the static radius of the spacetime governing a central supermassive black hole surrounded by the dark energy \cite{Stu:1983:BAC:,Stu-Hle:1999:PRD:}. On the other hand, the dark matter is relevant across the whole galaxy, and in the case of the polytrope models some instabilities can even predict existence of a central supermassive black hole \cite{Stu-etal:2017:JCAP:}. Clearly, we can expect a substantial role of dark matter around supermassive black holes in the galactic nuclei and these studies are of very high relevance in recent astrophysics, as they could introduce a variety of interesting phenomena that could be tested by recent sophisticated observational techniques. 

Some very rough estimates of the role of the dark matter can be realized in the framework of the very simplified model of dark matter shell introduced in \cite{Leu-Liu-Sue:1997:PRL:,Kon-Zhi:2011:RevModPhys:,Car-Pan:2017:NatAstr:,Kon-Zhi-Stu:2019:PRD:,Bar-Car-Pan:2017:PHYSR4:}. In fact, the studies of the influence on the black hole shadow give interesting predictions \cite{Kon:2019:PLB:} that were enforced by the studies of applicability of the geodesic models of QPOs \cite{Stu-Kot-Tor:2013:ASTRA:,Stu-etal:2020:Uni:} to explain the data observed in active galactic nuclei that are out of positive fitting for the vacuum spacetimes; it was demonstrated that even the simplest epicyclic resonance variant of the geodesic model enables data fitting in almost all the observed active galactic nuclei \cite{Stu-Vrb:2021:EPJPsub:}. 

Therefore, it is of very high importance to test the newly introduced exact "fluid-hairy" black hole solution \cite{Car-etal:2021:arxiv:} reflecting properly the influence of the realistic distribution of the dark matter -- this asymptotically flat solution with hair and regular horizon is found in the framework of Einstein's gravity coupled minimally to anisotropic fluid representing the dark matter. Its advantage is that the dark matter distribution follows the density profile of the Sersic type connected to the Herquist model
\begin{equation}
	\rho(r,a_0)= \frac{Ma_0}{2\pi r(r+a_0)^3}
	\label{rho}
\end{equation}
where $M$ denotes the total mass of the dark matter halo, and $a_0$ its characteristic lengthscale; this profile is observationally confirmed in elliptical galaxies \cite{Her:1990:ApJ:}, giving thus a strong support for consideration of its consequences in astrophysical phenomena. 

Properties of this hairy black hole spacetime has been recently studied in \cite{Car-etal:2021:arxiv:} for influence of the halo on the black hole shadow, innermost stable geodesic orbit, and gravitational quasinormal modes, while in \cite{Kon:2021:arxiv:} electromagnetic quasinormal modes are considered, along with the Unruh temperature. Here, we test the applicability of this fluid-hairy black hole spacetime for fitting the QPO data observed around supermassive black holes in active galactic nuclei, by using the geodesic model of QPOs. The goal is to solve an issue suggested in \cite{Smi-Tan-Wag:2021:ApJ:} that the geodesic model based on the frequencies of the orbital and epicyclic oscillatory motion in the vacuum black holes \cite{Stu-Kot-Tor:2013:ASTRA:} are not able to explain the observed data in active galactic nuclei, although this model is successful in the case of the microquasars \cite{Tor-etal:2011:ASTRA:,Stu-Kol:2016:ASTRA:}. We attempt to realize this program concentrating on the simple epicyclic resonance variant of the geodesic model of QPOs, applying this variant to all the sources (active galactic nuclei) considered in \cite{Smi-Tan-Wag:2021:ApJ:}. We thus determine the conditions restricting physically and astrophysically the parameters of the hairy black hole spacetimes, determine frequencies of the geodesic orbital and epicyclic motion in these spacetimes, and apply them in the epicyclic resonance variant of the geodesic model to the observed data. Finally, we give list of the hairy black hole spacetime parameters enabling fitting of the observational data in all the sources considered in \cite{Smi-Tan-Wag:2021:ApJ:}. 

Throughout the paper we use a space-like signature \mbox{$(-,+,+,+)$}, a system of units in which $G = c = 1$ and we restore them when we need to compare our results with observational data. Greek indices run from $0$ to $3$, Latin indices from $1$ to $3$.

\section{Fluid-hairy black holes}

The fluid-hairy black hole spacetimes describing a black hole immersed in a dark matter halo represented by an anisotropic fluid connected to the Einstein cluster model \cite{Ein:1939:AnnalsMath:,Ger-Ruf:2012:IJMPCS:}, constructed under assumption of density profile of the Hernquist type, has been recently constructed in \cite{Car-etal:2021:arxiv:}. Its line element is determined by the relations 

\begin{eqnarray}\label{e:met}
	\mathrm{d}s^2=&-&f\mathrm{d}t^2+\left(1-\frac{2m(r)}{r} \right)^{-1}\mathrm{d}r^2\nonumber\\
	&+&r^2\big(\mathrm{d}\theta^2 + \sin^2\theta \mathrm{d}\varphi^2\big),
\end{eqnarray}
where 
\begin{equation}
	m(r)= M+\frac{dM\, r^2}{\left(a_0+r\right)^2} \left( 1-\frac{2M}{r} \right)^2
	\label{e:mr}
\end{equation}
and
\begin{eqnarray}
	f&=& \left( 1-\frac{2M}{r} \right) \ee^\mathrm{T}, \nonumber \\
	\mathrm{T}&=& -\pi\sqrt{\frac{dM}{\xi}} + 2\sqrt{\frac{dM}{\xi}}\arctan{\frac{r+a_0-dM}{\sqrt{dM\ \xi}}}, \nonumber\\
	\xi &=& 2a_0-dM+4M, \nonumber \\
	\eta &=&  r+a_0-dM.
	\label{e:ff}
\end{eqnarray}
The Fluid-hairy black hole spacetime contains three parameters -- $M$ denotes the black hole mass, $dM$ denotes mass of the dark matter in the halo, $a_0$ denotes the length scale characterizing extension of the dark matter halo. For details of the stress energy tensor of the anisotropic dark matter and detailed derivation of this solution of the Einstein equations see \cite{Car-etal:2021:arxiv:}. We illustrate behavior of the mass function $m(r)$ determined by Eq.\ref{e:mr} in Fig.1, and behavior of the lapse function $f(r)$ determined by Eq.\ref{e:ff} in Fig.2. Both these functions are increasing with increasing radius of the mass configuration. 

The spacetime corresponds to the energy density distribution given by the relation \cite{Car-etal:2021:arxiv:} 
\begin{equation}
                   \rho = \frac{m'}{4\pi r^2} = \frac{2dM(a_0+2M)(1-2M/r)}{4\pi r(a_0 + r)^3}
\end{equation}
that is converted into the Hernquist formula at large radii. The fluid-hairy black hole horizon is located at $r_h=2M$, and its physical singularity is at $r=0$. However, for range of its parameters given as $dM > 2(a_0 + 2M)$, the Ricci and Kretschmann scalars diverge, giving therefore additional physical singularities, at radii $r_{sing}= dM - a_0 \pm \sqrt{dM^2 - 2dMa_0 - 4MdM}$. Of course, such singular case have to be considered as implausible from the point of view of astrophysics \cite{Car-etal:2021:arxiv:} -- these are thus not relevant for research of astrophysical phenomena. Moreover, in astrophysically natural conditions, we assume the inequalities $M << dM << a_0$. 

\section{Circular geodesics of the fluid-hairy black holes}
Motion of test particles is governed by the geodesics of the spacetime that are determined by the geodesic equation for the 4-momentum (wave vector) $p^{\mu}$ ($k^{\mu}$) of the massive (mass-less) particle. The geodesic equations  
\begin{equation}
\frac{\diff^2 x^\mu}{\diff \tau^2} + \Gamma^\mu_{\rho\sigma}\,\frac{\diff x^\rho}{\diff \tau} \frac{\diff x^\sigma}{\diff \tau}=0
\end{equation}
have to be accompanied by the norm condition for the geodesic four-velocity $u^{\mu}=dx^{\mu}/d\lambda$   
\begin{equation}
u^\mu u_\mu = g_{\mu\nu}\, u^\mu u^\nu = -\epsilon
\end{equation}
where the parameter $\epsilon = m^2$ for massive particles and $\epsilon = 0$ for photons and other mass-less particles; $\lambda$ is the affine parameter related to the proper time $\tau$ of massive particles as $\lambda = m\tau$. 

The spherical symmetry of the spacetime implies the geodesic motion fixed to central planes; the equatorial plane $\theta = \pi/2 = const$ can be selected for convenience, if only one particle is under consideration. The stationarity of the spacetime implies conservation of the covariant energy, axial symmetry implies conservation of the axial angular momentum 
\begin{equation}
    E = -p_t \qquad L = p_{\phi}.
    \label{e:conserv}
\end{equation}

We focus on the circular geodesics that are governing behavior of the Keplerian disks that appropriately describe the accretion phenomena under some of the plausible conditions in active galactic nuclei and microquasars \cite{Abr-Fra:2013:LRR:}. 

\begin{figure*}[ht] 
	\includegraphics[width=\linewidth]{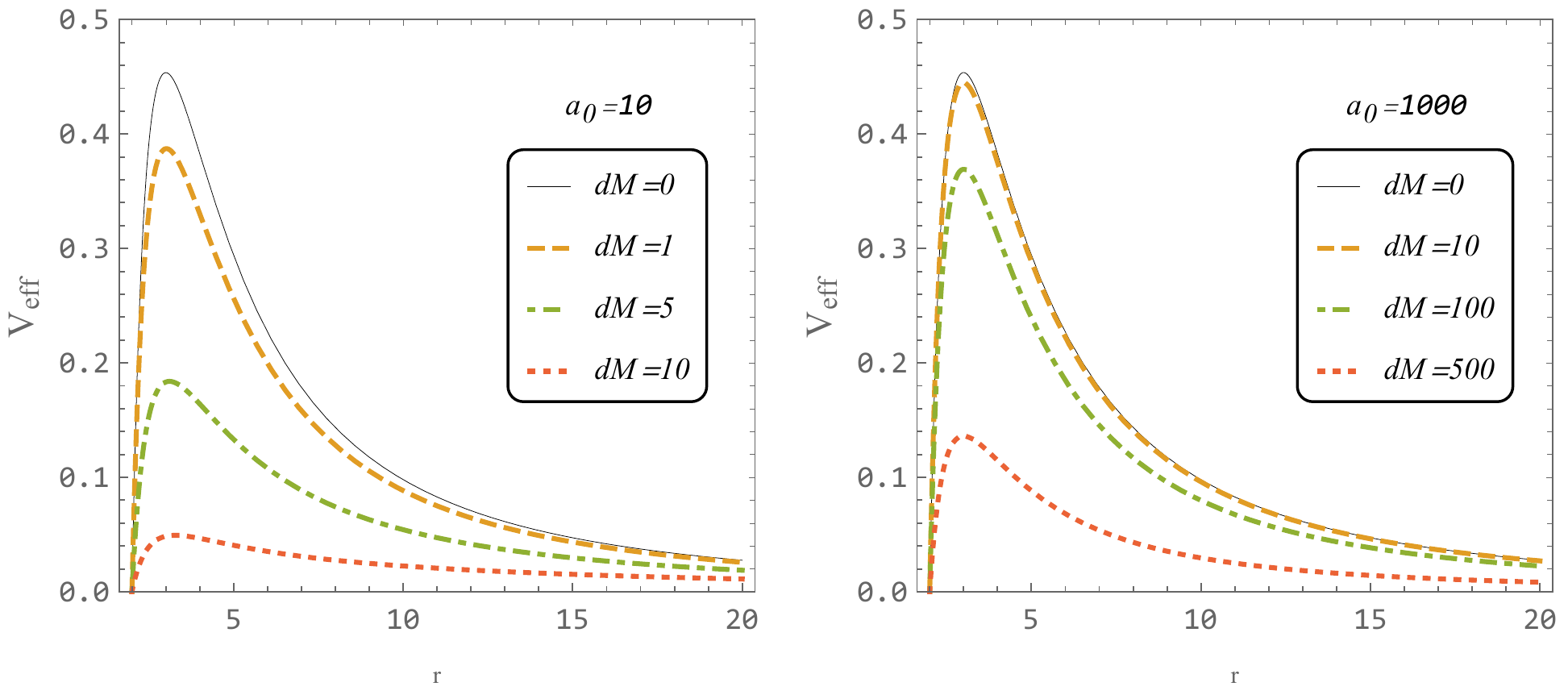}
	\caption{Effective potential of the massless test particles for various parameters $dM$ and $a_0$. Here we consider two case of $a_0=10,1000$, while the range of halo mass parameter takes values of $dM = 0, 1, 5, 10, 100, 500$. Recall that $M=1$.}
	\label{f:f2}
\end{figure*}

\begin{figure*} 
	\includegraphics[width=\linewidth]{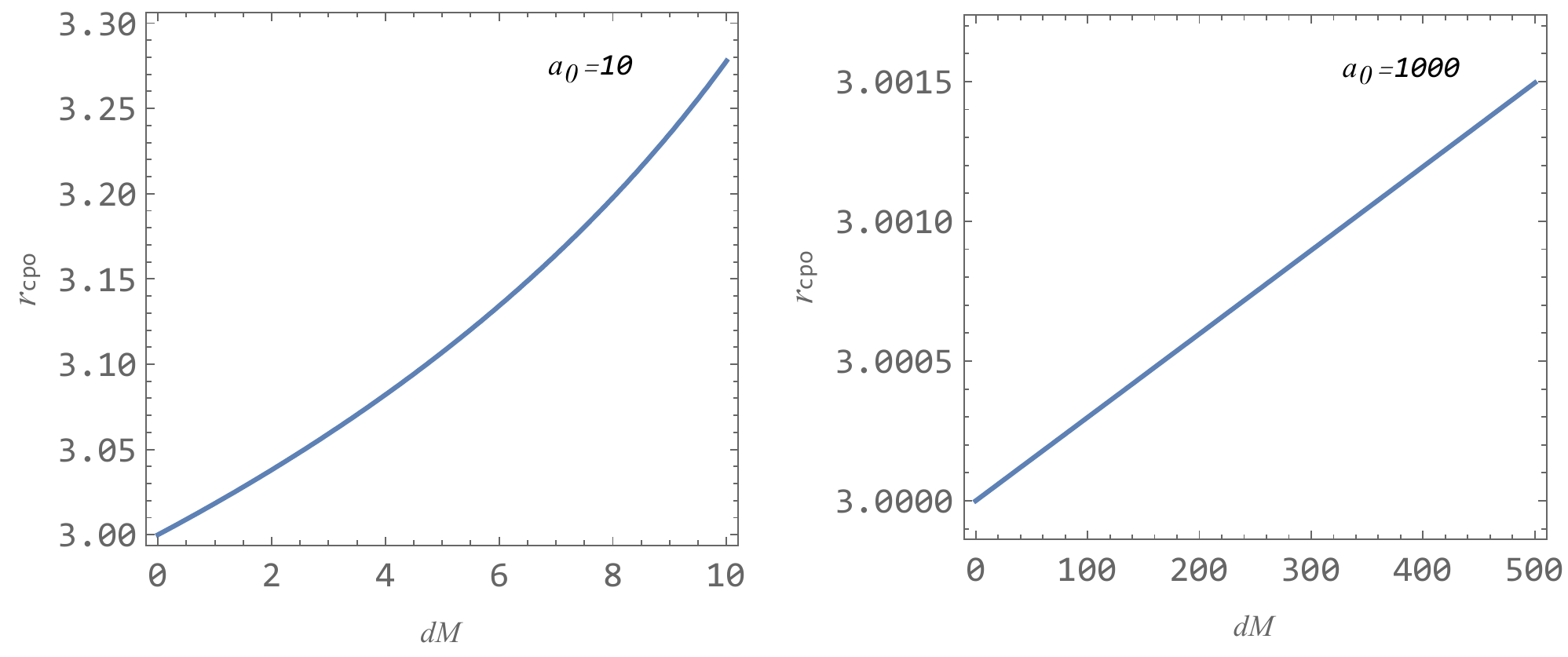}
	\caption{Dependence of radius of the circular photon orbit on parameter $dM$ for two fixed values of the extension parameter $a_0 = 10, 1000$.}
	\label{f:f1}
\end{figure*}

Geodesics of the mass-less particles (photons) are not dependent on the energy, being fully determined by the impact parameter $l=L/E$ (see e.g. \cite{Mis-Tho-Whe:1973:Gravitation:,Stu-Char-Sche:2018:EPJC:})  -- here we apply its inverse  
\begin{eqnarray}\label{e:const1}
b = \frac{E}{L} . 
\end{eqnarray}
The effective potential related to the inverse impact parameter then reads  
\begin{equation}
    V_\mathrm{eff}=f\frac{1}{g_{\theta\theta}(r)},
\end{equation}
where $f$ defined by (\ref{e:ff}) containing the mass function $m(r)$ defined by Eq.(\ref{e:mr}). The behavior of the photon effective potential is illustrated in Fig. \ref{f:f2} -- we can see qualitatively the standard type of the behavior as in the Schwarzschild spacetime. 
The circular geodesics of mass less particles are determined by local extrema of the effective potential so they have to satisfy the condition 
\begin{equation}
    \diff V_\mathrm{eff}/\diff r=\frac{L^2 \left(r\,f'-2 f\right)}{r^3} = 0 . 
\end{equation}
The circular null geodesics are thus given by the relation 
\begin{equation}
    rf'-2 f = 0 , 
\end{equation}
where 
\begin{eqnarray}
	\diff f/\diff r = f'= 2\frac{\ee^\mathrm{T}}{r^2}\left\{M + \frac{r\, dM (r-2 M)}{\eta^2+dM \xi}\right\}.
	\label{e:ff1}
\end{eqnarray}
The dependence of the radius of the photon circular orbit on the halo mass parameter $dM$ is for fixed values of the extension parameter $a_0$ illustrated in Fig. \ref{f:f1}. We can see that the dependence is strongest for the smaller extension parameter $a_0=10$, where in theconsidered interval of $dM$ the radius can increase by one tenth of the Schwarzschild value $r_{phS}=3$, while it is much lower for $a_0 = 1000$ where the increase is only by $\sim 1/1000$ of the Schwarzschild value, and the dependence is almost linear. 

All the circular null geodesics are unstable against radial perturbations. 

\FloatBarrier

For massive particle with rest mass $m$ we can introduce the specific energy and specific axial angular momentum as  
\begin{eqnarray}\label{e:const}
\mathcal{E} = \frac{E}{m_0}, \qquad
\cl = \frac{L}{m_0}.
\end{eqnarray}

In the equatorial plane ($\theta=\pi/2$) character of the radial motion of massive test particles can be governed by the effective potential in the form \cite{Mis-Tho-Whe:1973:Gravitation:} 
\begin{equation}
    V_\mathrm{eff}=f\left[\frac{\cl^2}{g_{\theta\theta}(r)}+1\right] . 
\end{equation}

The effective potential is illustrated in Figs. \ref{f:f5} - \ref{f:f7}. The character of the effective potential is similar to the Schwarzschild case in the fluid-hairy black hole spacetimes satisfying the physically realistic conditions (no additional physical singularities), so both stable and unstable circular geodesics are possible. Nevertheless, its quantitative influence on the properties of the circular geodesic motion could be significant. 
\begin{figure*} 
	\includegraphics[width=\linewidth]{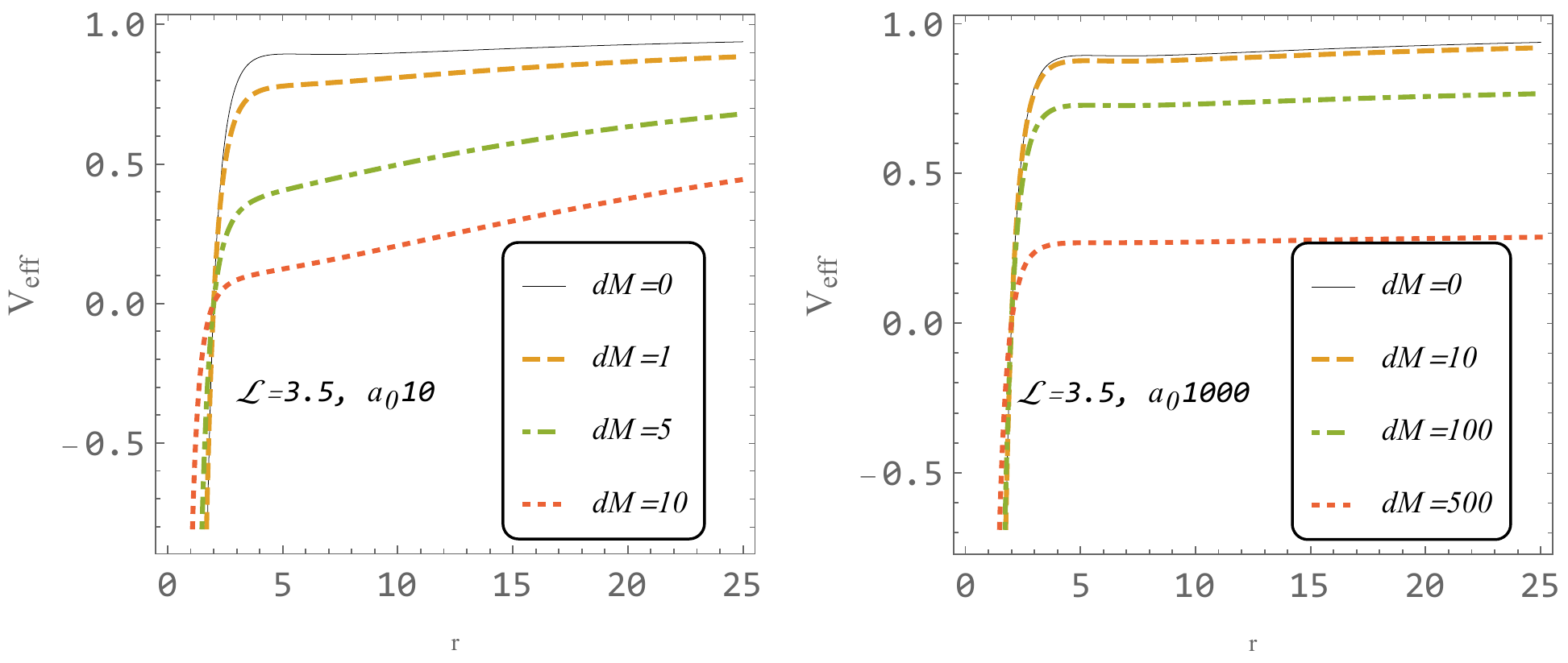}
	\caption{Effective potentials of material test particles for various parameters $\dm$, $\rs$, $\drs$ and fixed value of the specific angular momentum of the particle $\cl=3.5$. The bottom line represents resolution of the effective potential at low values of radius given for the same values of the parameters as in the upper line. This kind of magnification is applied in all the figures of the effective potential.}
	\label{f:f4}
\end{figure*}
\begin{figure*} 
	\includegraphics[width=\linewidth]{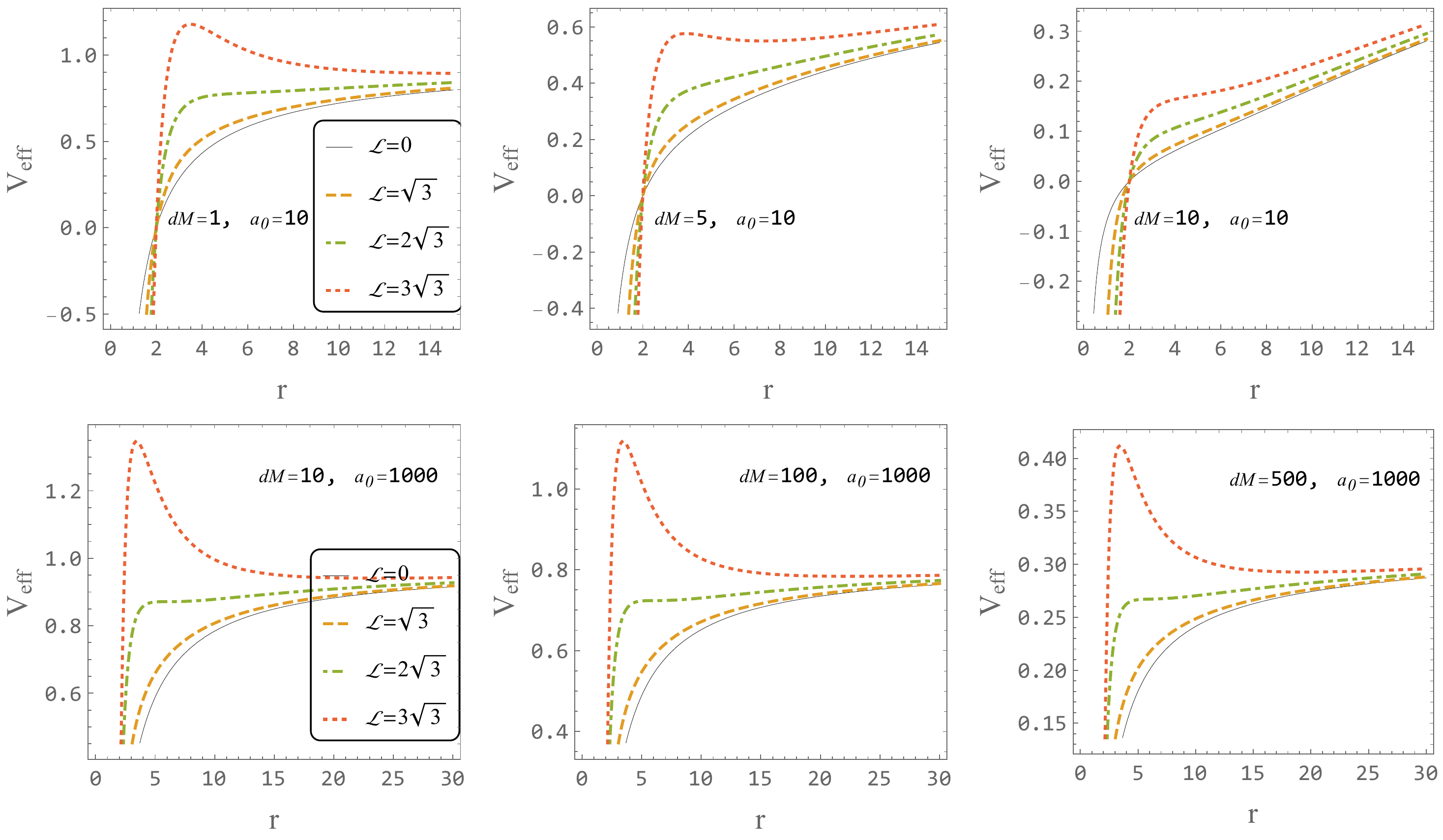}
	\caption{Effective potentials of material test particles for various values of $\cl$ and fixed value of $\dm$.}
	\label{f:f5}
\end{figure*}

The circular geodesics are given by the local extrema of the effective potential: 
\begin{equation}
	\frac{\diff V_{\mathrm{eff}}}{\diff r}=0,
\end{equation}
where
\begin{equation}
    \diff V_\mathrm{eff}/\diff r=\frac{r \left(\cl^2+r^2\right) f'-2 \cl^2 f}{r^3}.
\end{equation}
where $f'$ is determined by Eq. \ref{e:ff1}. We can thus conclude that the radial profile of the particle specific angular momentum $\cl$ reads 
\begin{equation}
    \cl_\mathrm{c}=\pm \frac{r^{3/2} \sqrt{f'}}{\sqrt{2 f-r f'}}.
    \label{e:lc}
\end{equation}
The specific energy of the particle following the circular geodesic is determined by the extreme point of the effective potential taken for the specific angular momentum of the particle and reads 
\begin{equation}
     \ce_{c} = f \left(\frac{r f'}{2 f-r f'}+1\right)
\end{equation}
Clearly, both radial profiles diverge at radii corresponding to the circular photon orbits. 

The specific angular momentum $\cl_{c}$ is well defined with both $+/-$ signs due to possible two equivalent orientations of the circular motion, but the specific energy $\ce_{c}$ has to be taken only with the positive sign as we accept only the so called positive root states (for details see \cite{Mis-Tho-Whe:1973:Gravitation:,Bic-Stu-Bal:1989:BAC:}). 

The innermost (marginally) stable circular geodesic (ISCO) is determined by the additional condition 
\begin{equation}
\diff^2 V_\mathrm{eff}(r_\mathrm{c})/\diff r^2 = 0,
\end{equation}
where
\begin{eqnarray}
    \diff^2 V_\mathrm{eff}/\diff r^2=\frac{r \left[r \left(\cl^2+r^2\right) f''-4 \cl^2 f'\right]+6 \cl^2 f}{r^4},
\end{eqnarray}
and 
\begin{eqnarray}
	&&\diff^2 f/\diff r^2 = f''= 4   \frac{\ee^\mathrm{T}}{r^3 \left[\eta^2+dM \xi\right]^2}\nonumber \\
	&&\left\{dM r^2  (2 M-r)\eta+2 dM\, M r \left[\eta^2+dM \xi\right]+ 
	r^2 dM^2 (r-2 M)-M  \left[\eta^2+dM \xi\right]^2\right\}.
	\label{e:ff2}
\end{eqnarray}

The condition for ISCO is determined by the relation 
$2 r f'^2 - f \left(3 f'+r f''\right)=0$.
The behavior of the ISCO radius, given by numerical calculations, is represented in Fig. \ref{f:f6}. Now we can see that the halo mass $dM$ causes significant shift in the ISCO position for both considered values of $a_0$ taht in the end of the considered intervals of $dM$ causes shift down to the black hole horizon for more than $1/10$ of the Schwarzschild value of $r_{ISCO}=6$.  

\begin{figure*} 
	\includegraphics[width=\linewidth]{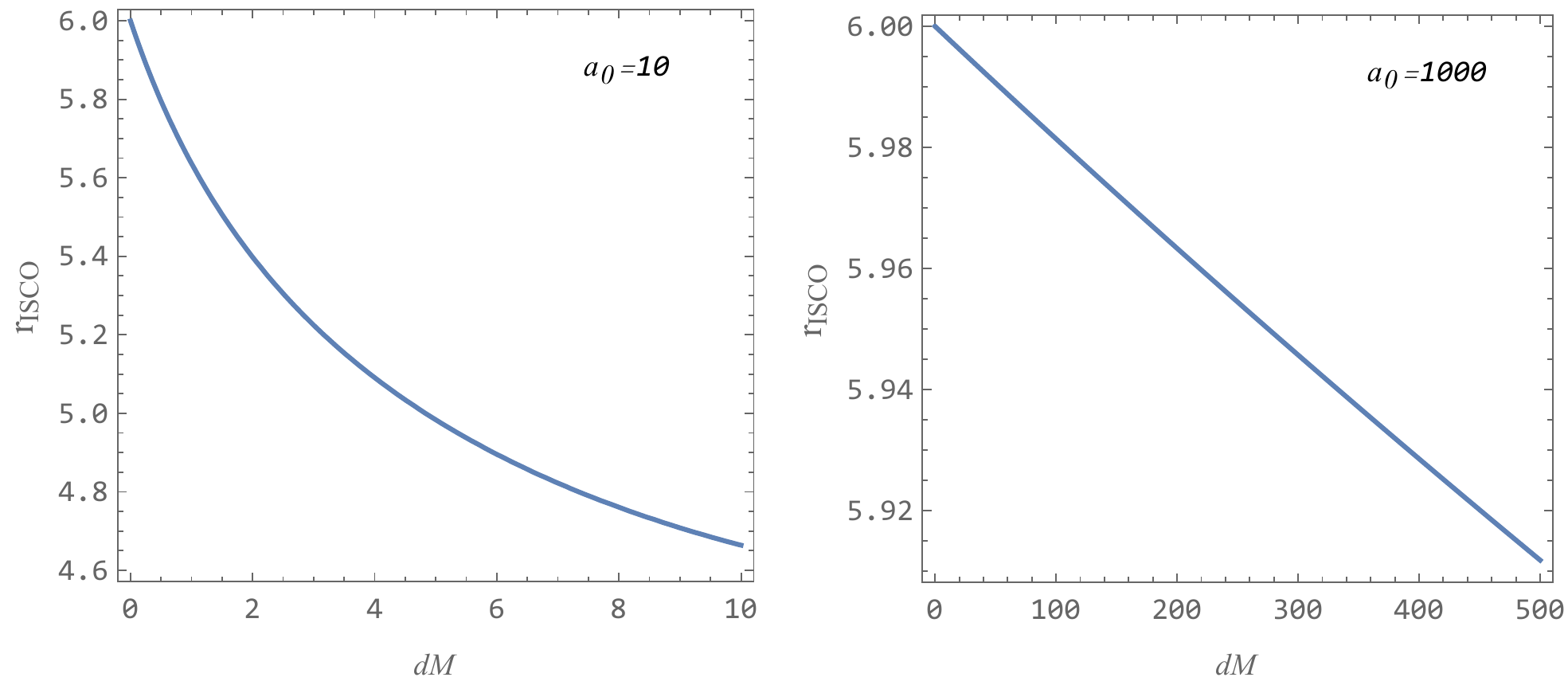}
	\caption{The radial profile of the ISCO function giving the innermost stable circular orbit for various values of the parameters $dM$ and $a_0$.}
	\label{f:f6}
\end{figure*}
\FloatBarrier

\section{Frequencies of the epicyclic orbital motion}
The HF QPOs generated in Keplerian accretion disks orbiting black holes represent a very efficient possibility to study the strong-field limit of gravity given by general relativity or its alternatives \cite{Bambi:2018:ADP:}\footnote{Of course, sufficiently weak magnetic fields could have also extremely important role in modeling HF QPOs, if the black hole is magnetized and the orbiting matter is slightly charged \cite{Kol-Stu-Tur:2015:CLAQG:,Kol-Tur-Stu:2017:EPJC:,Stu-etal:2020:Uni:,Tur-etal:2020:ApJ:}. Of course, strong magnetic fields could significantly overcome the influence of the black hole gravity, even in close vicinity to the horizon \cite{Stu-Kol:2016:EPJC:,Tur-etal:2020:ApJ:,Kol-Tur-Stu:2021:PHYSR4:,Stu-Kol-Tur:2021:Unisub:}.}. The black holes with accretion disks are expected in the microquasars with a stellar-mass black hole (see \cite{McCli-Rem:2006:BHbinaries:,Tor-etal:2011:ASTRA:} 
and in active galactic nuclei with a supermassive black hole - see \cite{Smi-Tan-Wag:2021:ApJ:} for an overview.  

However, we cannot exclude some alternative objects to black holes as the source of strong gravity in the microquasars and active galactic nuclei that could lie behind the observed HF QPOs. As such alternatives can be considered the Kerr naked singularities and superspinars \cite{Stu:1980:BAC:,Tor-Stu:2005:ASTRA:,Stu-Sche:2012:CLAQG:,Stu-Sche:2013:CLAQG:,Kot-etal:2017:ASTRA:}, or wormholes \cite{Stu-Vrb:2021:Universe:}. Here we continue the study of the possible role of the dark matter concentrated around accreting black holes on the character of the HF QPOs generated in the Keplerian disks and fit the theoretical predictions to the observational data. As in our previous study directed to the simplest model of a Schwarzschild black hole surrounded by the dark matter shell influencing the accreting matter only gravitationally \cite{Stu-Vrb:2021:EPJPsub:}, we use the epicyclic resonance model that was demonstrated to be a good representative of the class of the geodesic models of HF QPOs \cite{Stu-Kot-Tor:2013:ASTRA:} in fitting the observational data.

Let us consider a slightly perturbed particle following in the equatorial plane $\theta=\pi/2$ a stable circular orbit at $r_c$ that undergoes epicyclic oscillations that will be considered in the linear regime with displacements $r=r_\mathrm{c}+\delta r$ and $\theta=\pi/2+\delta\theta$. The epicyclic motion in the radial and latitudinal directions around the stable position at $r_c$ is equivalent to the harmonic oscillation governed by the motion equations taking the form 
\begin{eqnarray}
\delta\ddot{ r}+\bar{\omega}_r^2 \delta r = 0, \ \ \ 
\delta\ddot{ \theta}+\bar{\omega}_\theta^2 \delta\theta = 0.
\end{eqnarray}

The frequencies of the epicyclic oscillatory motion can be determined by the Hamilton formalism applied in \cite{Kol-Stu-Tur:2015:CLAQG:,Stu-Kol:2016:EPJC:,Kol-Tur-Stu:2017:EPJC:} -- for an alternative perturbative approach see \cite{Ali-Gal:1981:GRG:,Tur-Stu-Kol:2016:PHYSR4:}. For completeness, we give also the frequency of the orbital motion, i.e., the azimuthal (Keplerian) frequency $\omega_\phi = \omega_K$. 

First, the radial and latitudinal epicyclic frequencies, and the azimuthal frequency, $\bar{\omega_r}$, $\bar{\omega_\theta}$, $\bar{\omega_{\phi}}$, are given beig related to the local observer at $r_c$. The Hamiltonian generally reads 
\begin{eqnarray}\label{e:ham}
H=\frac{1}{2}g^{\alpha\beta}p_\alpha p_\beta+\frac{m^2}{2} . 
\end{eqnarray}
In our study it can be split into its dynamic and potential parts
\begin{eqnarray}
H=H_{\textrm{dyn}}+H_{\textrm{pot}},
\end{eqnarray}
where
\begin{eqnarray}
H_{\textrm{dyn}}&=& \frac{1}{2}\Big(g^{rr}p_r^2 +g^{\theta\theta}p_\theta^2\Big),\\
H_{\textrm{pot}}&=& \frac{1}{2}\Big(g^{tt}\mathcal{E}^2+g^{\phi\phi}\cl^2+1\Big).\label{e:hpot}
\end{eqnarray}
The potential part of the Hamiltonian represents the effective potential implying the radial and latitudinal epicyclic frequencies $\bar{\omega}_r$ and $\bar{\omega}_\theta$. The orbital (azimuthal) frequency $\bar{\omega}_\phi$ can be obtained directly from Eq. (\ref{e:conserv}) and radial profiles of the specific energy and specific angular momentum of the circular orbit. The resulting formulas read 
\begin{eqnarray}
\label{e:bom}
\bar{\omega}_\phi &=& \frac{\cl_\mathrm{c}}{g_{\theta\theta}}, \\ \nonumber 
\bar{\omega}_r^2 &=& \frac{1}{g_{rr}}\frac{\partial^2 H_{\textrm{pot}}}{\partial r^2},\\ \nonumber
\bar{\omega}_\theta^2 &=& \frac{1}{g_{\theta\theta}}\frac{\partial^2 H_{\textrm{pot}}}{\partial \theta^2}.
\end{eqnarray}
Applying (\ref{e:met}), (\ref{e:lc}), (\ref{e:const}), (\ref{e:hpot}) in (\ref{e:bom}), we arrive to the locally measured frequencies in the form 
\begin{eqnarray}
\label{e:bomegas}
\bar{\omega}^2_\theta &=& \bar{\omega}^2_\phi = \frac{ f'}{r\left(2 f-r f'\right)},\\ \nonumber
\bar{\omega}^2_r &=& \frac{\left[2 r f'^2 - f \left(3 f'+r f''\right)\right] \left[r-2 m(r)\right]}{r^2 f \left(r f'-2 f\right)}.
\end{eqnarray}

For the observational purposes, the epicyclic frequencies and the orbital frequency have to be related to distant observers. For this reason we have to make transformation of the form (\ref{e:bomegas}) related to the local observer to those corresponding to the frequencies measured by the static observers at large distance (formally at infinity). Moreover, we have to use now the standard (SI) units (Hz). For these purposes, we apply on (\ref{e:bomegas}) the scaling by the redshift factor between the orbiting particle and the observers at infinity, and obtain the frequencies that can be applied in fitting the observational data: 
\begin{eqnarray}\label{e:trfreq}
\nu_{i}=\frac{1}{2\pi}\frac{c^3}{G\,M}\frac{\bar{\omega_{i}}}{-g^{tt}\mathcal{E}},
\end{eqnarray}
where $i={\phi, r, \theta}$. For $dM = 0$ we arrive to the know results corresponding to the epicyclic motion in the Schwarzschild spacetime \cite{Kat-Fuk-Min:1998:BOOK:,Tor-Stu:2005:ASTRA:}. 
\begin{figure*} 
	\includegraphics[width=\linewidth]{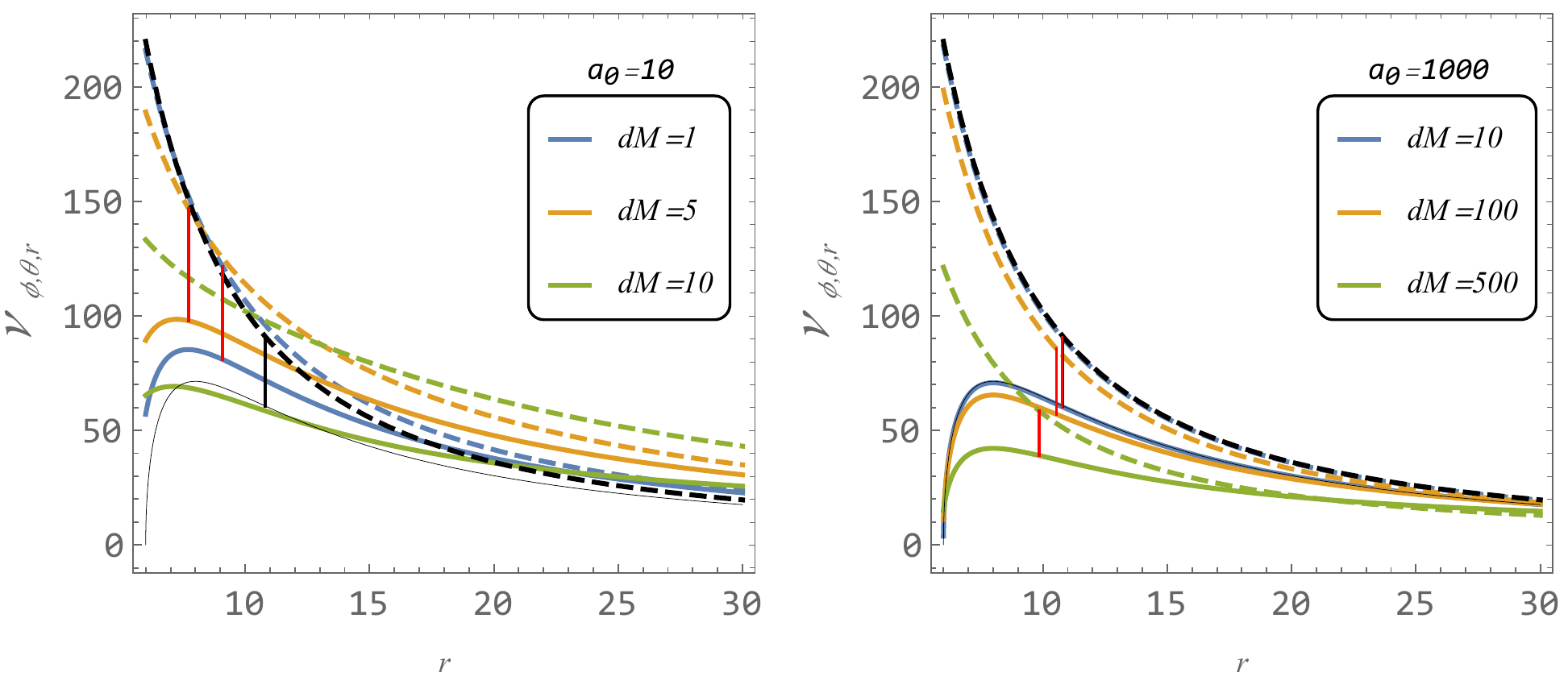}
	\caption{The radial profiles of the radial epicyclic frequency $\nu_r$ (colored solid lines) and the orbital frequency $\nu_\phi$ that is equal to the latitudinal epicyclic frequency $\nu_\theta$ (both colored dashed) of test particles in spacetime with dark matter shell are given for various parameter $a_0$ and  $dM$ in units $M$ and $M=10\, M_\odot$. Black are epicyclic frequency $\nu_r$ (solid) and $\nu_\phi$ with $\nu_\theta$ (both dashed) in the Schwarzschild spacetime. Vertical black line indicate the position of frequency ratio 3:2 for Schwarzschild case and red vertical lines for our spacetime.}
	\label{f:f7}
\end{figure*}
\FloatBarrier

The radial profiles of the orbital and epicyclic frequencies are illustrated for typical values of the parameters $dM$ and $a_0$ in Fig. \ref{f:f7}. The figure clearly demonstrates that the character of the frequency profiles strongly depends on both the considered parameters. 

For large values of the extension parameter (the case of $a_0 = 1000$), the modifications due to changes of the halo mass parameter behave usually regularly; with increasing value of $dM$ the radial profiles of all the frequencies are decreasing, but we observe no intersections of them -- at the frequency ratio $\nu_\theta:\nu_r=3:2$, i.e., the crucial value for the observed twin HF QPOs in both microquasars and active galactic nuclei \cite{Tor-Abr-etal:2005:AA:,Stu-Kot-Tor:2013:ASTRA:}, the largest value of the upper frequency corresponds to the lowest value of $dM$. However, we can see that the latitudinal frequency profile corresponding to the largest value of $dM=500$ intersects the radial frequency profiles corresponding to all the chosen values of $dM$; the crossing points shifts to the ISCO radius with decreasing values of $dM$. The analysis shown that the frequency ratio $\nu_\theta:\nu_r=3:2$ can be found only at one radius, contrary the case of magnetized black holes with orbiting charged particles \cite{Stu-etal:2020:Uni:}. 

For low values of the extension parameter (the case of $a_0 = 10$) behavior of the frequency profiles is more complex -- the profiles corresponding to the lowest value of $dM=1$ are mainly located between the profiles corresponding to $dM=5$, giving the highest profiles, and $dM=10$, giving the lowest profiles. But the profiles have also some intersection points under which the statement of the previous sentence is not holding. The latitudinal frequency profile can intersect the radial frequency profiles
Nevertheless, again the frequency ratio $\nu_\theta:\nu_r=3:2$ is possible only at one radius.

\section{Epicyclic resonance model applied for HF QPOs observed in microquasars and around supermassive black holes}

Now we consider the HF QPOs in order to estimate the parameters of the black holes (or their alternatives) in the accretion systems in  various sources, i.e., in microquasars corresponding to binary systems with a stellar mass black hole, or in active galactic nuclei with supermassive black holes. The HF QPOs are observed with frequencies of hundreds of Hz in microquasars, and by frequencies by even 10 orders lower in the active galactic nuclei containing extremely supermassive black holes (their mass could exceed $10^{10}M_{\odot}$). It is well known that the observed frequencies demonstrate inverse-mass scaling that is characteristic for the epicyclic frequencies of the orbital motion \cite{Rem-McCli:2005:ARAA:}. For this reason are the so called geodesic models of HF QPOs, based on combinations of the frequencies of the orbital motion and related epicyclic oscillatory motion, promising and widely applied in modeling of the observed HF QPOs. Of special interest is the fact that the HF QPOs are usually observed in the rational ratio \cite{McCli-Rem:2006:BHbinaries:}, indicating action of resonant phenomena -- namely, the often observed frequency ratio $3:2$ \cite{Tor-etal:2011:ASTRA:} can be explained by the parametric resonant phenomena \cite{Klu-Abr:2001:ACTAASTR:,Tor-Abr-etal:2005:AA:}.
Description of the large number of variant of the geodesic model can be found in \cite{Stu-Kot-Tor:2013:ASTRA:}, for inclusion of an electromagnetic interaction between a magnetized black hole and slightly charged accretion matter see \cite{Stu-Kol:2016:EPJC:,Kol-Tur-Stu:2017:EPJC:,Stu-etal:2020:Uni:,Tur-etal:2020:ApJ:}. 

In all the variants of the geodesic model, the upper $\nu_\mathrm{UP}$ and lower $\nu_\mathrm{l}$ observed frequencies are explained by a combination of the orbital and epicyclic frequencies (note that these frequencies of the geodesic motion are relevant also for oscillations of slender tori \cite{Rez-Yos-Zan:2003:MNRAS:}). 

As first, the relativistic precession model was introduced \cite{Ste-Vie:1999:ApJ:} where identification of the frequencies is given by $\nu_\mathrm{UP} = \nu_{\phi}=\nu_\mathrm{K}$ and $\nu_\mathrm{l} = \nu_\mathrm{K} - \nu_\mathrm{r}$. For estimates in our paper, we use the simple epicyclic resonance model introduced in \cite{Tor-Abr-etal:2005:AA:} using the identifications $\nu_\mathrm{UP} = \nu_\theta$ and $\nu_\mathrm{l} = \nu_\mathrm{r}$. 

We demonstrate the role of the fluid-hairy black hole spacetime parameters characterizing the dark matter halo of the Hernquist type for fitting to observational data corresponding to the observed $3:2$ twin HF QPOs in microquasars \cite{McCli-Rem:2006:BHbinaries:}, and in HF QPOs observed around supermassive black holes in active galactic nuclei \cite{Smi-Tan-Wag:2021:ApJ:}, expecting to solve the issue of inability of the variant of the geodesic model to fit data obtained from the surrounding of the supermassive black holes. 

We have to stress that in the case of black holes surrounded by a dark matter shell, the above defined identification of the upper and lower frequencies allows also the equivalent ratio $2:3$ caused by the special character of the radial profiles of the epicyclic frequencies that occurs in the spacetimes with a special selection of the dark matter shell. We thus test, if the dark matter halos enable to fit the data for the supermassive black holes, and to what extend they can keep the satisfactory fits obtained by the geodesic model in the case of microquasars. 

\begin{figure*} 
	\includegraphics[width=\linewidth]{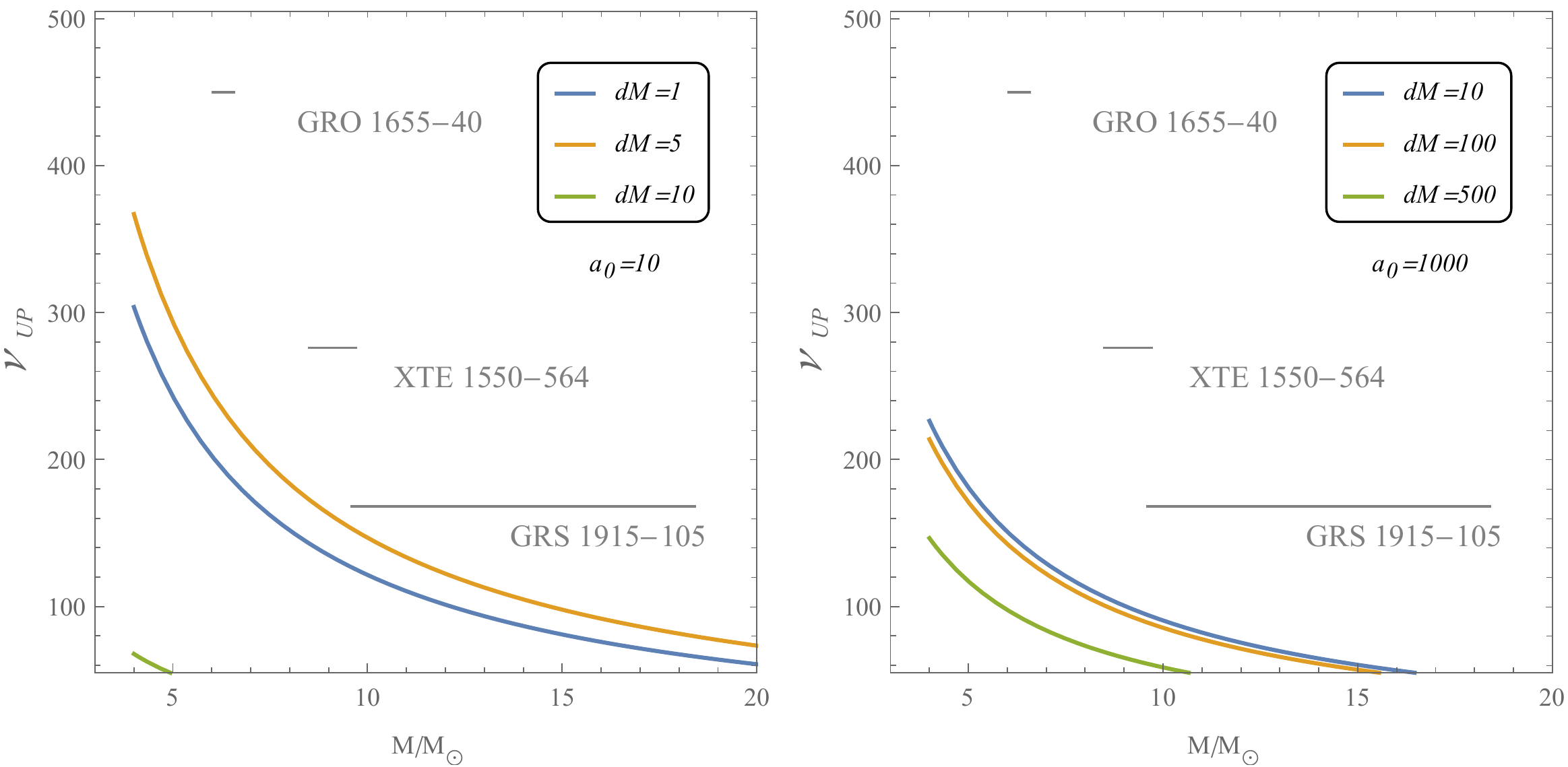}
	\caption{The dependence of the upper frequency on the mass of the central object compared to three observed microquasars. The solid curves describe $\nu_\theta:\nu_r=3:2$, for various parameters $a_0$ and $dM$.}
	\label{f:f8}
\end{figure*}

\begin{figure*} 
	\includegraphics[width=\linewidth]{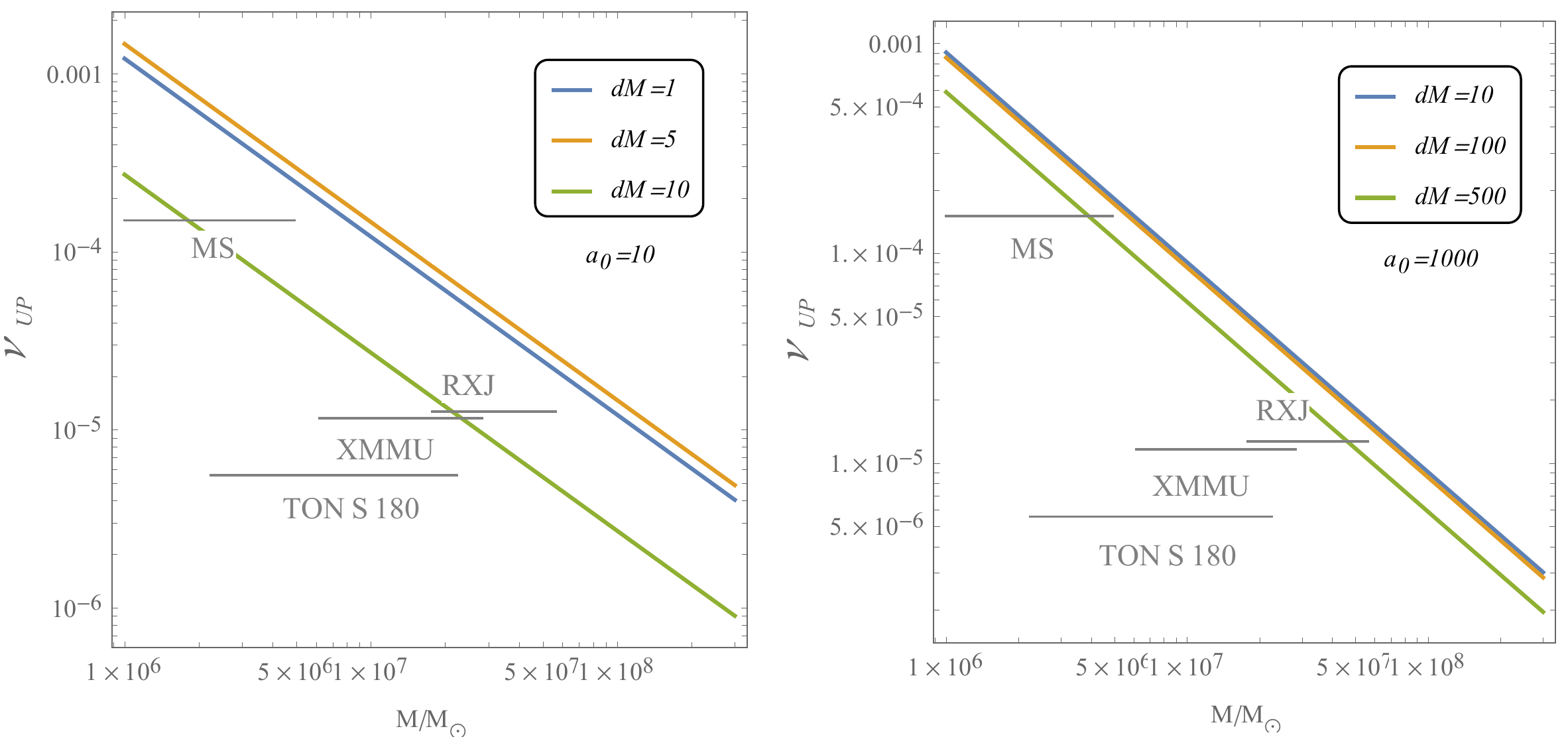}
	\caption{The dependence of the upper frequency on the mass of the central object compared to three observed quasars \cite{Smi-Tan-Wag:2021:ApJ:}. The solid curves describe $\nu_\theta:\nu_r=3:2$ for various parameters $a_0$ and $dM$.}
	\label{f:f9}
\end{figure*}

\begin{figure*} 
	\includegraphics[width=\linewidth]{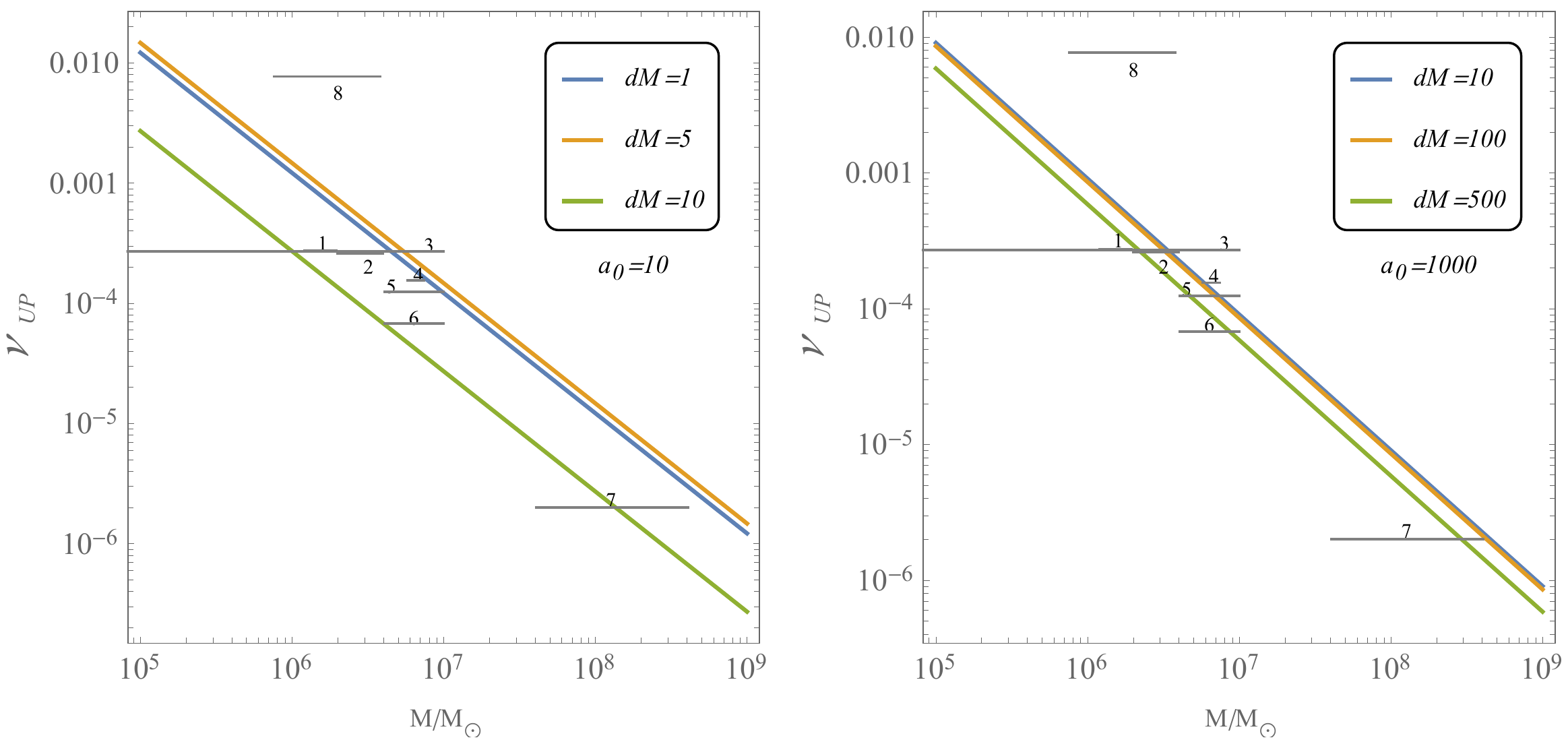}
	\caption{The dependence of the upper frequency on the mass of the central object compared to three observed quasars \cite{Smi-Tan-Wag:2021:ApJ:}. The solid curves describe $\nu_\theta:\nu_r=3:2$ for various parameters $a_0$ and $dM$.}
	\label{f:f10}
\end{figure*}

In the fitting procedure, we apply the method developed in \cite{Stu-Kot-Tor:2013:ASTRA:}, concentrating on the most promising and simple case of the epicyclic resonance variant of the geodesic model. We consider all the sources introduced in \cite{Smi-Tan-Wag:2021:ApJ:} -- the results of the fitting procedure are presented in Fig \ref{f:f8} for microquasars, and in \mbox{Figs \ref{f:f9} and \ref{f:f10}} for the case of supermassive black holes assumed in active galactic nuclei. 

As the fluid-hairy spacetimes are spherically symmetric, and rotation is not included into the spacetime structure, our results could be taken as illustrations of the possible role of the dark matter halo in the HF QPO phenomenon. for detiled fitting a rotational generalization of the fluid-haimy spacetime is necessary. However, our results still give a crucial information on the possible role of the dark matter halo, demonstrating the possitive effect of the halo on possibility to fir the data in active galactic nuclei, an spoiling the fits in microquasars. Such results thus imply the important consequence, namely, that the role of the dark matter halo can be only insignificant around stellar mass black holes, but it could be quite relevant around supermassive black holes in the galaxy centers. Of course, such results could be expected intuitively. 

The rotational terms added to the fluid-hairy geometry related to the dark matter Hernquist halo would surely modify the results of the fitting procedure making it more precise, but it is clear that the basic positive trends predicted in the fitting by the spherically symmetric metric survive, and the limiting restrictions on the mass and extension of the halo obtained in the simple spherical model will be only slightly modified due to the rotation effects, similarly to modifications observed in comparison of the fitting due to the Schwarzschild and Kerr metric -- see e.g. \cite{Stu-Kol:2016:ASTRA:}. The final results of restrictions on the amount of dark matter contained in vicinity of supermassive black holes at all the considered sources due to the fitting procedure in the fluid-hairy spherical geometry, summarized for all the considered sources in Table \ref{t:tab2}, can be taken seriously, if the basic trends are expected. 

Because of lacking of the black hole rotation, we are not considering the errors in the frequency measurements, and give only rough numbers of the metric parameters enabling the fitting; we expect a more detailed study related to some selected sources under consideration of the rotational variant of the black hole surrounded by dark matter shell. 

In our fitting procedures and giving our estimates we assumed two values of $a_0=10$ and $a_0=1000$, and gave for the considered sources the range of the parameter $dM$ allowing the successful fitting. Usually two ranges of allowed values of the parameters were found, one for the lower values of $dM\leq10$, the other for higher values of $10<dM<500$. Details are presented in Table \ref{t:tab2}; in case of three sources, namely ASASSN-14li, Sw J164449.31+573451, and TON S 180, the fitting procedure was not successful. For the source Sw J164449.31+573451 the observed frequency seems to be quite off its possible relation to the geodesic model. We have checked that fitting by the relativistic precession model gives similar result and the three problematic sources remain problematic, as increasing the parameter $a_0$ means decreasing of ability to fit the data in the active galactic nuclei consideration.

Results presented in our paper could be considered as a possible explanation of the QPO phenomena observed around supermassive black holes, but inversely, they can be considered as a rough information about limits on amount of dark matter around supermassive black hole in the considered sources, and extension of the region of its highest concentration, giving thus an important restriction of the dark matter in the central regions of active galactic nuclei.


\begin{table*}[h]
\caption{Observations of QPOs around Supermassive Black Holes \cite{Smi-Tan-Wag:2021:ApJ:}}
\label{t:tab1}
\begin{tabular*}{\textwidth}{@{\extracolsep{\fill}}cllllllll@{}}
\hline
Number  &  Name & BH spin & log $M_\mathrm{BH}$ [$M_\odot$] & $\nu_\mathrm{UP}$ [Hz]  \\
\hline
1  &  MCG-06-30-15	&	$>0.917$	&	$6.20^{+0.09}_{-0.12}$	&	$2.73\times10^{-4}$	\\
2  &  1H0707-495	&	$>0.976$	&	$6.36^{+0.24}_{-0.06}$	&	$2.6\times10^{-4}$	\\ 
3  &  RE J1034+396	&	0.998	&	$6.0^{+1.0}_{-3.49}$	&	$2.7\times10^{-4}$	\\
4  &  Mrk 766	&	$>0.92$	&	$6.82^{+0.05}_{-0.06}$	&	$1.55\times10^{-4}$	\\
5  &  ESO 113-G010	&	0.998	&	$6.85^{+0.15}_{-0.24}$	&	$1.24\times10^{-4}$	\\
6  &  ESO 113-G010b	&	0.998	&	$6.85^{+0.15}_{-0.24}$	&	$6.79\times10^{-5}$	\\
7  &  1H0419-577	&	$>0.98$	&	$8.11^{+0.50}_{-0.50}$	&	$2.0\times10^{-6}$	\\
8  &  ASASSN-14li	&	$>0.7$	&	$6.23^{+0.35}_{-0.35}$	&	$7.7\times10^{-3}$	\\
-  &  TON S 180	&	$< 0.4$	&	$6.85^{+0.5}_{-0.5}$	&	$5.56\times10^{-6}$\\
-  &  RXJ 0437.4-4711	&	-	&	$7.77^{+0.5}_{-0.5}$	&	$1.27\times10^{-5}$	\\
-  &  XMMU J134736.6+173403	&	-	&	$6.99^{+0.46}_{-0.20}$	&	$1.16\times10^{-5}$	\\
-  &  MS 2254.9-3712	&	-	&	$6.6^{+0.39}_{-0.60}$	&	$1.5\times10^{-4}$ \\
-  &  Sw J164449.3+573451	&	-	&	$7.0^{+0.30}_{-0.35}$	&	$5.01\times10^{-3}$		\\
\hline
\end{tabular*}
\end{table*}

\begin{table}[]
\centering
\caption{Best fits of $a_0=\{10,100\}$ and $dM$ parameters on observed Supermassive Black Holes with QPO signature}
\label{t:tab2}
\begin{tabular}{ccc}
\hline
\multicolumn{1}{|c|}{\multirow{2}{*}{Name}} & \multicolumn{2}{c|}{$dM$}                                       \\ \cline{2-3} 
\multicolumn{1}{|c|}{}                      & \multicolumn{1}{c|}{$a_0=10$} & \multicolumn{1}{c|}{$a_0=1000$} \\ \hline
(1) MCG-06-30-1  & $10$ & - \\
(2) 1H0707-495 & - & $10-500$ \\
(3) RE J1034+396  & $(1-10)$  & $10-500$ \\
(4) Mrk 766 & $1$ & $10-100$ \\
(5) ESO 113-G010a & $1$ & $10-500$ \\
(6) ESO 113-G010b & $10$ & $500$ \\
(7) 1H0419-577 & $10$ & $100-500$ \\
ASASSN-14li (8) & - & - \\
TON S 180 & - & - \\
RXJ 0437.4-4711 & $10$ & $10$ \\
XMMU J134736.6+173403 & $10$ & - \\
MS 2254.9-3712 & $10$ & $10$ \\
Sw J164449.3+573451 & - & - \\ \hline
\end{tabular}
\end{table}
\FloatBarrier

\section{Conclusions}
We studied frequencies of the orbital and related epicyclic oscillatory motion around stable circular geodesics in the field of fluid-hairy black holes surrounded by dark matter halo of the Hernquist type. The epicyclic frequencies were applied in the epicyclic resonance variant of the geodesic model of the twin HF QPOs \cite{Stu-Kot-Tor:2013:ASTRA:}.

We have demonstrated that in the case of the microquasars the influence of the spacetime parameters $a_0$ and $dM$ tends to decrease the possibility to obtain satisfactory fits, but in the case of the frequencies observer in active galactic nuclei, the parameters act in positive way, and some combinations of the spacetime parameters enable satisfactory fits for most of the considered sources with supermassive black holes \cite{Smi-Tan-Wag:2021:ApJ:}. Moreover, our results give strong constraints on the amount of the dark matter located around stellar mass black holes, and give relevant estimates on the amount of dark matter in the Hernquist consenrated around supermassive black holes in active galactic nuclei.

Of course, extensions of our basic study are necessary that should consider the role of the black hole rotation, and the way of estimation of the mass of the central black holes that could introduce a necessity of slight modification of the fitting procedure applied in the present paper. 

\acknowledgments

The authors acknowledge the institutional support of the Research Centre for Theoretical Physics and Astrophysics, Institute of Physics, Silesian University in Opava. ZS acknowledges the Czech Science Foundation grant No. 19-03950S. 

\bibliographystyle{jhep}
\bibliography{references}

\end{document}